\documentclass[useAMS,usenatbib]{emulateapj}

\def\simless{{\th \rlap{\raise 0.5ex\hbox{$\scriptstyle  {<}$}}
    {\lower 0.3ex\hbox{$\scriptstyle  {\sim}$}} \th }}  
\def\simgreat{{\th \rlap{\raise 0.5ex\hbox{$\scriptstyle  {>}$}}
    {\lower 0.3ex\hbox{$\scriptstyle  {\sim}$}} \th }}  
\def\greateq{{\th \rlap{\raise 0.5ex\hbox{$\scriptstyle  {>}$}}
    {\lower 0.3ex\hbox{$\scriptstyle  {-}$}} \th }}  
\def\lesseq{{\th \rlap{\raise 0.5ex\hbox{$\scriptstyle  {<}$}}
    {\lower 0.3ex\hbox{$\scriptstyle  {-}$}} \th }}  
\def\th{\thinspace}
\def\ts{{\raise 0.3ex\hbox{$\scriptstyle {\th \sim \th }$}}}

\usepackage{epsfig}

\newcommand{\nudot}{\dot{\nu}}
\def\hide#1{}

\begin{document}

\shorttitle{The origin of timing noise}

\title{An alternative interpretation of the timing noise in accreting millisecond pulsars}
\author{Alessandro Patruno \altaffilmark{1}, Rudy Wijnands \altaffilmark{1}, Michiel van der Klis \altaffilmark{1} }
\altaffiltext{1}{Astronomical Institute ``Anton Pannekoek,'' University of
Amsterdam, Kruislaan 403, 1098 SJ Amsterdam, Netherlands}


\begin{abstract}
The measurement of the spin frequency in accreting millisecond X-ray
pulsars (AMXPs) is strongly affected by the presence of an unmodeled
component in the pulse arrival times called 'timing noise'.  We show
that it is possible to attribute much of this timing noise to a pulse
phase offset that varies in correlation with X-ray flux, such that
noise in flux translates into timing noise.  This could explain many
of the pulse frequency variations previously interpreted in terms of
true spin up or spin down, and would bias measured spin frequencies.
Spin frequencies improved under this hypothesis are reported for six
AMXPs.  The effect would most easily be accounted for by an accretion
rate dependent hot spot location.

\end{abstract}

\section{Introduction}

Precise orbits and spin parameters have now been reported in 9
accreting millisecond X-ray pulsars (AMXPs; see
\citealt{w04,pou06,d07} for reviews, \citealt{pat09} for the AMXP most
recently found).  Yet, controversy still surrounds the interpretation
of the observed pulse time-of-arrival (TOA) records.  The reason for
this is the presence of a red ``timing noise'' component in the TOA
residuals.  While on the time scales of hours relevant to determining
the orbit this noise has only a moderate effect, its amplitude is
large on the timescales of weeks to months required to measure the
pulse frequency $\nu$ and its time derivative $\nudot$. The origin of
the timing noise is unknown, there is no satisfactory model for it,
and no agreement on how to deal with it when measuring $\nu$ and
$\nudot$.  Some authors fit a constant $\nu$ (\citealt{h08}, H08 from
now on), while others fit a constant $\nudot$ or more complex models,
interpreted as due to accretion torques (\citealt{f05,b06,pap07,c08,
r08}).  Both methods leave unmodeled timing noise in the residuals,
and in determining orbital and spin parameters arbitrary rejection is performed of data segments, or of pulse
profile harmonics, showing 'too much' noise (see Patruno et al. 2009b).

In two AMXPs it was already noted that on short ($<$10 d) timescales
X-ray flux and TOA residuals correlate (J1814, \citealt{pap07}) or
anticorrelate (J1807, \citealt{r08}), and on the timescales of weeks
similar to the duration of an AMXP outburst these correlations were
found to be stronger for a constant $\nu$ model than for constant
$\dot{\nu}$ (\citealt{w08}, Patruno et al. 2009b).  

Here we show that such correlations are common in AMXPs and can be
interpreted in the sense that a given flux level induces a given,
constant, TOA offset, so that trends in flux bias the measured $\nu$
and $\dot\nu$ values.  Our findings then suggest that the timing noise
is not dominated by accretion torques but instead by accretion rate
dependent variations in hot spot location on the neutron star surface.

\section{Observations and data reduction}

We use all {{\it RXTE} PCA} public data for 6 AMXPs (Table 1). We
did not analyze HETE J1900.4$-$2455 and SAX J1748.9$-$2021 as their
weak and intermittent pulsations require special analysis, nor SWIFT
J1756.9$-$2508 and Aql X-1, whose pulse episodes were too brief to be
useful.

We refer to \citet{j06} for PCA characteristics and RXTE absolute
timing.  We used all available Event and GoodXenon data, rebinned to
1/8192 s and in the 2.5--16 keV band that maximizes S/N.  We folded
512-s data chunks, keeping only those with S/N$>$3--3.3$\sigma$,
giving $<$1 false pulse detection per source.  We detect both a
fundamental ($\nu$) and a first overtone ($2\nu$) in our pulse
profiles of J1808, J1807, and J1814 and only a fundamental in J1751,
J00291, J0929. The former three sources show strong pulse shape
variability, so that the fiducial point defining the pulse TOA becomes
ill defined (cf. Patruno et al. 2009b).  Therefore, we measured the
TOAs of fundamental and overtone separately, and then separately
fitted them with a Keplerian orbit plus a linear and possibly a
parabolic term representing $\nu$ and $\dot{\nu}$.  The first three
sources also have strong timing noise and without modeling this noise
both models give reduced $\chi^{2}$$\gg$1, the latter three have
weaker timing noise and reduced $\chi^{2}$ closer to 1, but both
models remain statistically unsatisfactory.

\begin{figure*}[t]
  \begin{center}
\hbox{
    \rotatebox{-90}{\includegraphics[width=0.48\columnwidth]{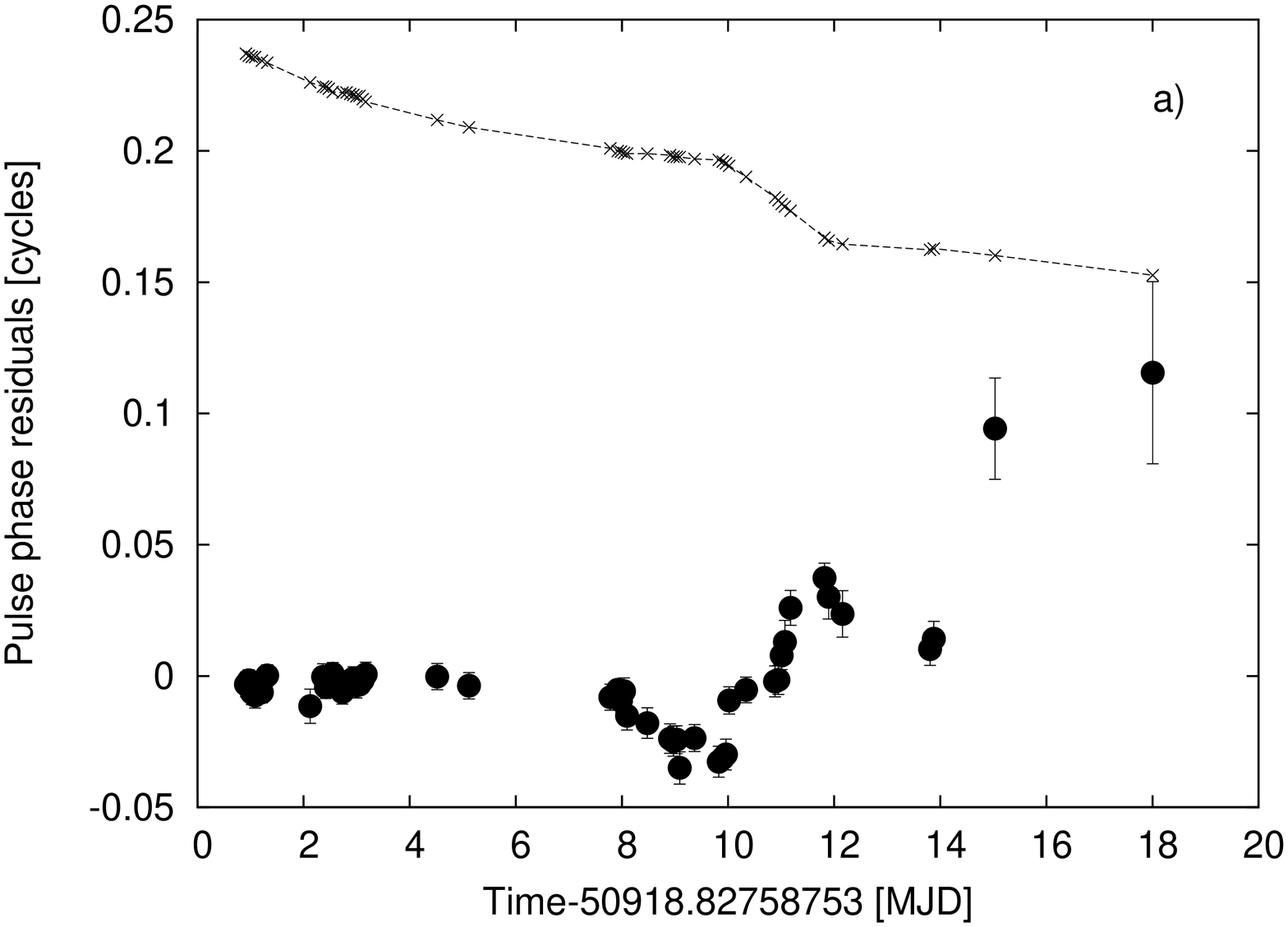}}
    \rotatebox{-90}{\includegraphics[width=0.48\columnwidth]{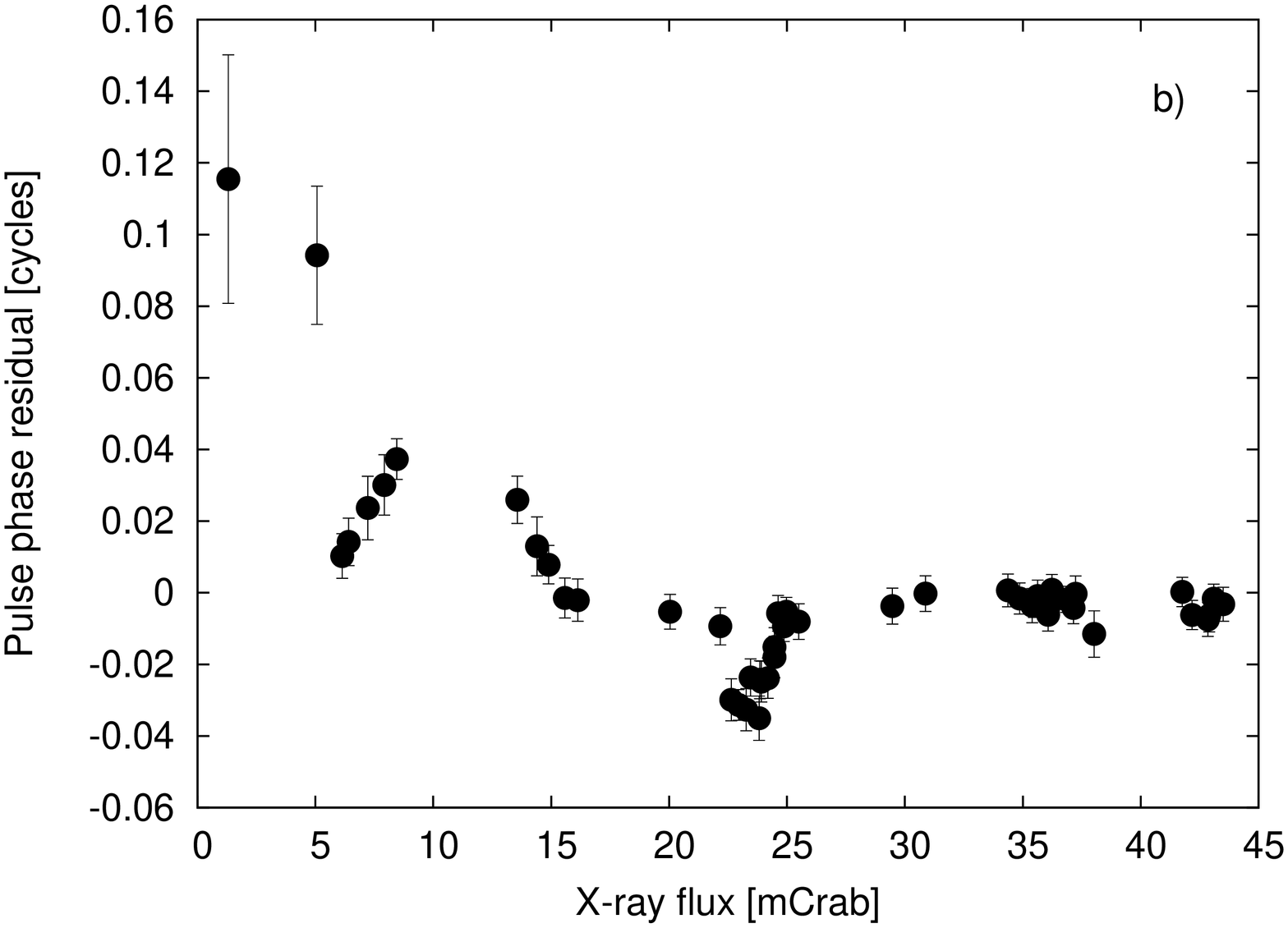}}
    \rotatebox{-90}{\includegraphics[width=0.48\columnwidth]{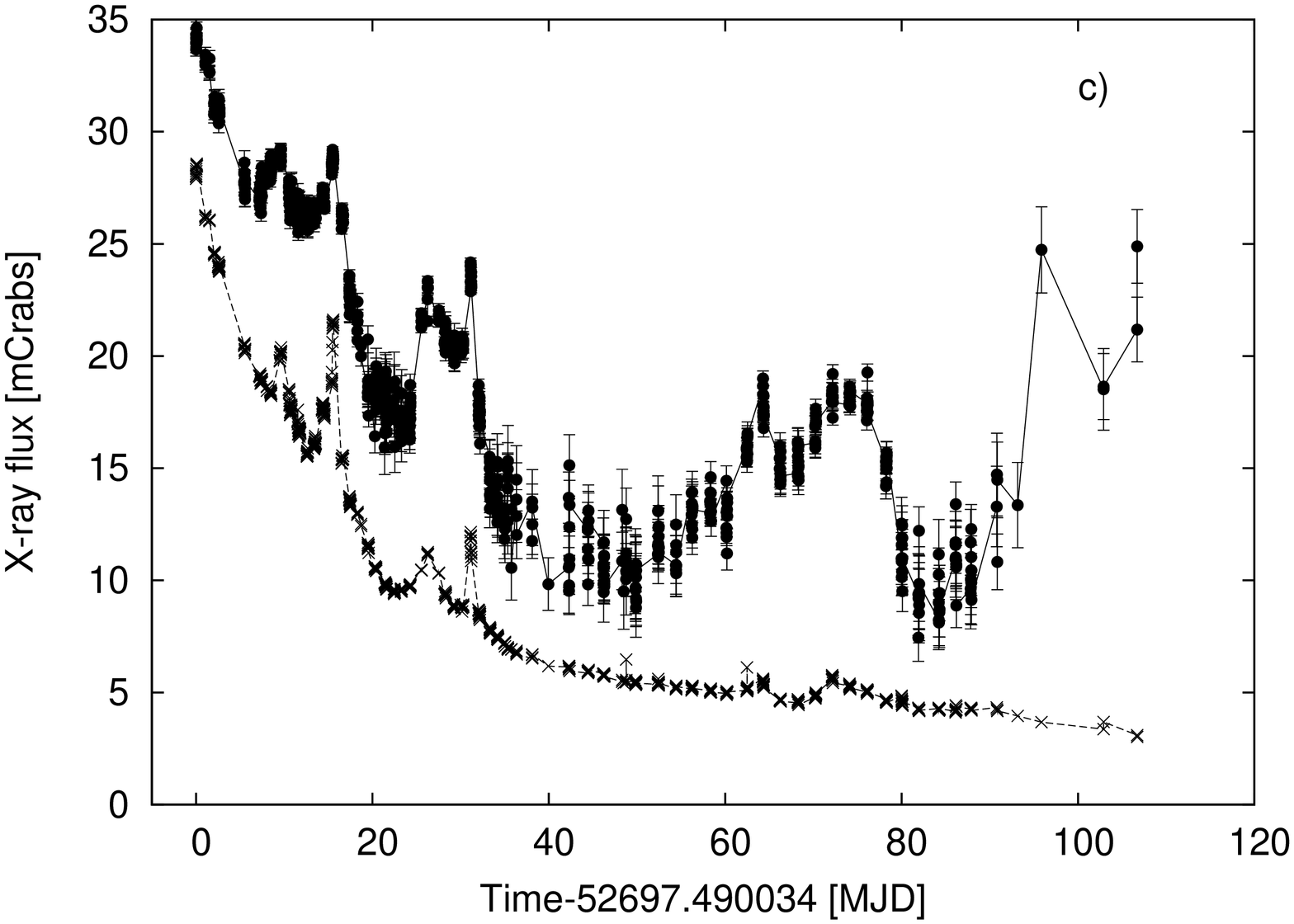}}}
  \end{center}
\caption{{\bf a.} Phase residuals ($\Delta t \times\nu$) from a
standard fit of a constant frequency model to the fundamental
frequency TOAs in the 1998 outburst of J1808 ({\it bullets}) and
simultaneous 2.5--16 keV light curve ({\it crosses}) in arbitrary
units.  Short term correlations are clearly present, but the long term
trend in flux is not seen in phase. {\bf b.} Phase vs. flux for the
same data.  {\bf c.} X-ray light curve and phase residuals of
J1807. The phase residuals are shown upside down and in arbitrary
units to better visualize the correlation of the short term phase
fluctuations with the X-ray flux.}
\label{fig:one}
\end{figure*}


\begin{figure*}[th!]
  \begin{center}
    \rotatebox{0}{\includegraphics[width=1.0\textwidth]{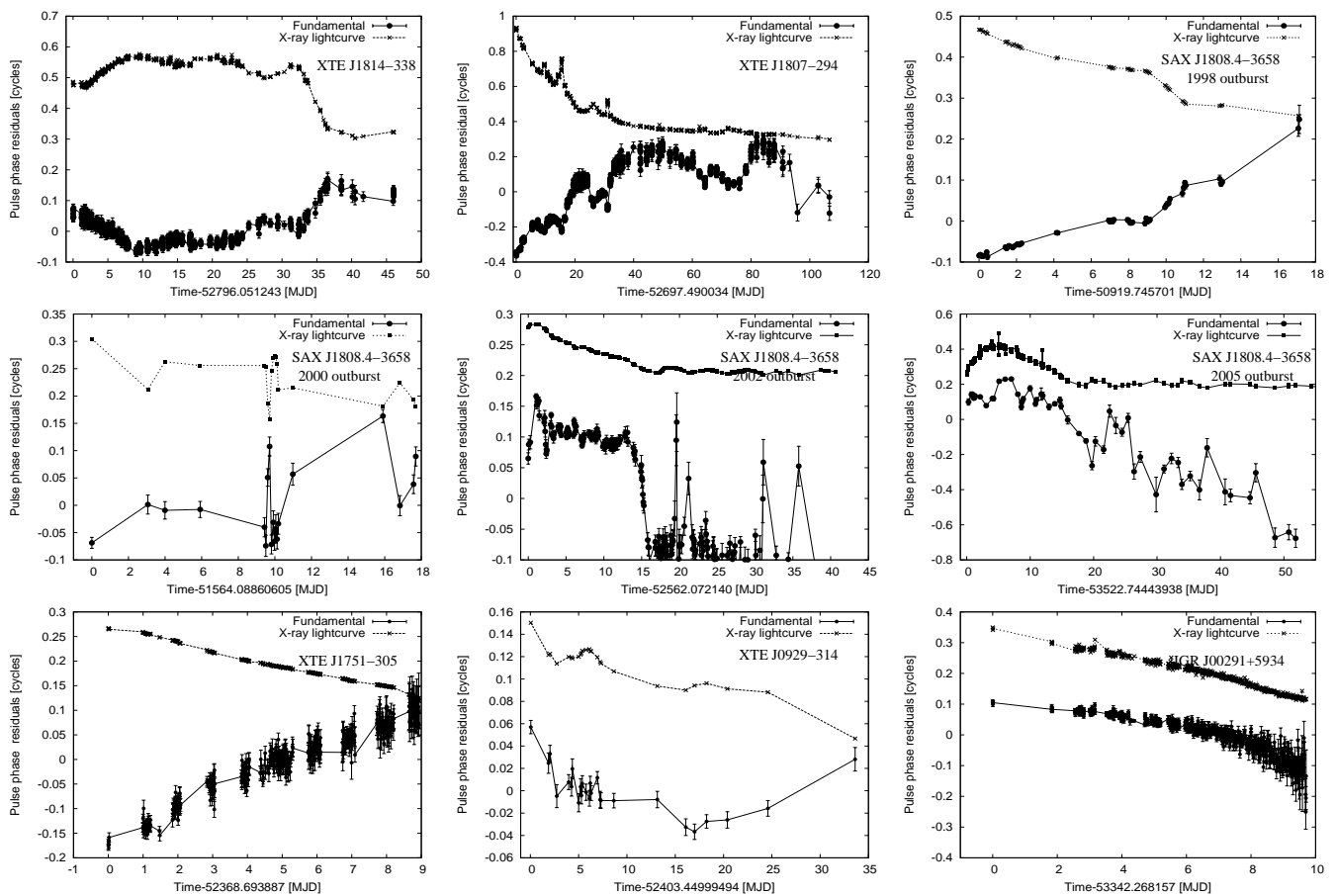}}
  \end{center}
\caption{Phase residuals of the fundamental frequency and 2.5--16 keV
light curves; as Fig.~\ref{fig:one}a, but with $\nu$ that optimizes the
phase-flux correlations, see text.  Phase-flux correlations can be
seen on all timescales. \label{fig:two}}
\end{figure*}

\section{Phase-flux correlations}

Figure~\ref{fig:one}a (bullets) shows the TOA residuals $\Delta t$
from a standard fit of a constant $\nu$ model, expressed as a phase
residual in units of pulse cycles: $\Delta\phi\equiv\nu\Delta t$.  The
phases show structures that are clearly anti-correlated with short
term flux variations (top trace; the concavity at day 2-10, the
bump around day 10, the slower decay after day 12).  However, there is
no correlation with the long term flux decay. Indeed, plotting the
phases against flux (Fig.~\ref{fig:one}b) no correlation is seen.

The reference (ephemeris) pulse frequency $\nu$ selected by a standard
$\chi^2$ fit is the one that distributes the residuals evenly among
positive and negative values; this choice is arbitrary and other
choices, introducing a net slope in the plot of residuals vs. time,
are equally valid (cf. Patruno et al. 2009b).  With this in mind we now
investigate the hypothesis that not only the short-term variations in
flux correlate with phase, but also the long term trend in flux
correlates with a similar trend in phase over the entire outburst. If
true, the best fit pulse frequency is not exactly the spin frequency,
but contains a bias that can in principle be removed.

\begin{figure*}[th!]
  \begin{center}
    \rotatebox{0}{\includegraphics[width=1.0\textwidth]{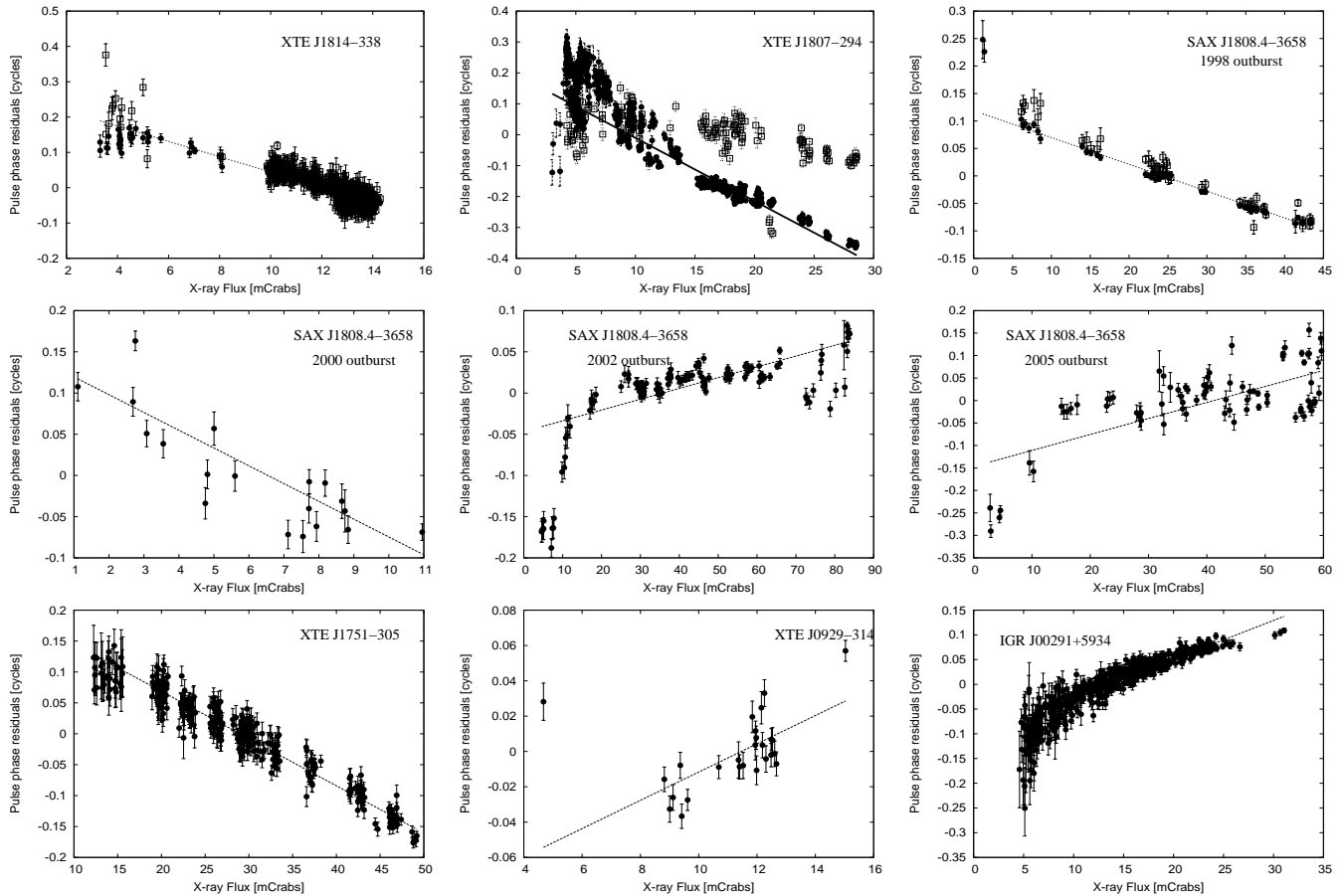}}
  \end{center}
\caption{Phase-flux correlations for fundamental ({\it black circles}) 
and overtone ({\it gray squares}). Best linear fits to the fundamental
phases are shown; deviations mostly occur at low flux.  All phase
residuals are TOA residual multiplied by pulse frequency, so overtone
and fundamental scales match.
\label{fig:three}}
\end{figure*}

Figure~\ref{fig:two} provides similar plots as Fig.~\ref{fig:one}a for
each outburst, but now choosing $\nu$ such as to maximize the
phase-flux correlation by minimizing the $\chi^2$ of a linear fit to
phase vs. flux (Fig.~\ref{fig:three}).  The difference with the
standard technique can be seen by comparing Fig.~\ref{fig:one}a with
the top right panel of Fig.~\ref{fig:two}.  Carefully scrutinizing
Figs.~\ref{fig:two} and \ref{fig:three}, it is clear that short term
correlations and anticorrelations such as in Fig.~\ref{fig:one}a are
ubiquitous in our sample, and that with few exceptions (discussed
below) with a proper choice of $\nu$ (Table 1) these can be matched
with a correlation on long timescales. 
In J1814, the slope of the short term correlation (measured
by first subtracting best fit parabolae from flux and phase records)
is -16.3$\pm$0.5 cycle/mCrab, and that of the long term one -18.9$\pm$0.3 
cycle/mCrab. So, it looks like it is {\it phase} that correlates to
flux here, not its second derivative $\dot{\nu}$ as in standard
accretion torque models. If we remove this X-ray flux effect from the
 TOAs of J1814 and then fit a $\nudot$ model, $\nudot$ is a factor 15 
lower than using the standard method.

The phase-flux correlation we find is positive for some sources and
negative for others, and in different outbursts of J1808 both signs
occur, but always has the same sign on both long and short time scales.
The full phase range implied by the best correlation to the flux
decay is always less than one cycle.  In J1814 and in the '98 outburst
of J1808 the correlation is the same for fundamental and overtone,
while in J1807 for the overtone it has a slope $\approx0.25$ that of
the fundamental. In the '00, '02 and '05 outbursts of J1808, the
phase of the fundamental correlates with flux whereas the 1st overtone
exhibits a weak anti-correlation. However, since the 1st overtone has a
low S/N and short term fluctuations are difficult to detect, we
decided to exclude the 1st overtone from our analysis.

In addition, during the J1808 \emph{re-flaring state} (\citealt{w04})
at the end of the '02 and '05 outbursts (but not the -- sparsely
sampled -- '00 one) the correlations break down: other modes of
accretion from the disk might play a role here (see \citealt{pat08}).
We excluded these re-flaring states when reporting $\nu$ in Table
1. In Fig. 3 we plot only the points included in the fit; the lines
correspond to the frequencies reported in Table 1.  In J1807, J1808
and J0929 the pulse phases deviate from the linear
correlation below some flux treshold.

The $\chi^2$ values in Table 1 are still unsatisfactory, however, they
are a statistically highly significant factor $2-5$ smaller in J1807,
J1814 and in the 1998 outburst of J1808, and similar in the other
six outbursts, to the $\chi^2$ one obtains from standard methods when fitting
a $\nudot$ model. So, accounting for a dependence of phase on flux by
a simple uniform linear relation produces a significantly
better fit than a $\nudot$ model for 3 of our 9 outbursts. In
Fig. 1c  we show light curve and inverted phase residuals of J1807
together; clearly there is a good correlation on all timescales, but
the relation is not linear. Indeed, Fig.~\ref{fig:three} suggests
that still better results might be obtained allowing for the more
complex (curved or broken) phase-flux relations seen there, but that
is beyond scope of this Letter.

\begin{table*}
\caption{Inferred spin frequencies}
\scriptsize
\begin{tabular}{lccccc}
\hline
\hline
Source name & Short & Outburst date [yr]& $\nu$ (Fund., Hz)& $\nu$ (1st overt., Hz)& $\chi^{2}/\rm dof$ \\
XTE J1807$-$294 & J1807 & 2003 & 190.62350712(3) & 190.62350712(3)& 12806.1/765 (F), 1724.5/146 (1st)\\
XTE J1814$-$338 & J1814 & 2003 & 314.35610872(3) & 314.35610874(3)& 1348/602 (F),  1086.6/555 (1st)\\
SAX J1808.4$-$3658$^{a}$ & J1808 & 1998,2000,2002,2005  & 0.52(3),0.31(3),0.21(2),0.06(2) &  0.50(5)& 114/47 (F), 93.6/45 (1st)\\
XTE J0929$-$314 & J0929 & 2002 & 185.10525437(2) & & 422.1/206\\
XTE J1751$-$305 & J1751 & 2002 & 435.31799405(5) & &441.6/306\\
IGR J00291$+$5934 &  J00291 & 2004  & 598.89213048(3)& & 1251.9/521\\
\hline%
\end{tabular}
\mbox{
\tablecomments{All the uncertainities quoted correspond to
$\Delta\chi^{2}=1$. F=fundamental frequency; 1st=1st overtone\\ $^{a}$
The frequencies of J1808 are relative to an offset frequency of
400.975210 Hz. The 1st overtone can be measured only for the 1998
outburst. The $\chi^{2}/\rm dof$ refers to the 1998 outburst.\\}}
\label{tab:}
\begin{tabular}{l}
\\
\end{tabular}

\end{table*}

In J1808, $\nu$ decreases between outbursts; a linear fit to the four
mean pulse frequencies gave $\dot{\nu}=-0.56\pm 0.20\times
10^{-15}\rm\,Hz/s$ with $\chi^{2}/\rm dof$=9.7/2 (H08).  Our new fit
has $\dot{\nu}=-1.9\pm0.2\times
10^{-15}\rm\,Hz/s$, approximately 3 times that measured by H08.  Our
$\chi^{2}/\rm dof$=6.4/2, reducing to 1.6/2 including the
astrometric uncertainity (cf.  eq. A1, A2 in H08).  While other than
H08 we use only the fundamental, whose phase-flux correlation is
always detected, the different $\dot\nu$ comes from removing the flux
bias from the phases.

\section{Discussion}

Short term ($<$10 d) correlations or anticorrelations, depending on
outburst, between pulse phase and X-ray flux are ubiquitous in our
AMXPs, and a considerable fraction, up to $\approx 97\%$ of the
variance of the timing noise, can be explained from the X-ray flux
variability if we assume that phase depends directly on flux.  These
correlations can be extended smoothly and maintaining sign to the
longest accessible time scales (weeks), i.e., to include a direct
correlation of phase with flux as it decays in each outburst, by a
proper choice of pulse reference frequency.  This strongly suggests
that there is a direct physical link between instantaneous flux level
and phase that is very different from the correlation between X-ray
flux and spin frequency derivative predicted by standard accretion
theory (e.g., \citealt{b97}), and that the pulse frequency derivatives
measured in AMXPs are (mostly) not the direct result of accretion
torques.  Of course, the phase-flux correlation we observe might still
(and in fact is plausible to) arise through a common parameter, i.e.,
accretion rate $\dot M$.

The fact that the observed range of phase residuals is less, and
usually much less, than one cycle over each outburst allows an
interpretation where a given flux level induces a given constant phase
offset by affecting the location of the hot spot on the neutron star
surface.  Hot spot motion was recently discussed for AMXPs in various
contexts.  It was observed in numerical simulations \citep{r03,r04},
and it was noted that for hot spots near the rotation pole small
linear shifts can cause large pulse phase shifts \citep{l08}.  If the
accretion disk inner radius is related with the X-ray flux, then at
different flux levels the accreting gas will attach to different
magnetic field lines and hence fall on different locations on the
surface, producing a hot spot that moves in correlation with flux (and
might also change in shape and size, cf. H08, \citealt{l08}). The hot
spot motion will bias the average frequency over the outburst if the
hot spot longitude gradually varies from a different value at the
beginning of the outburst to the end, and can mimic the effect of a
spin frequency derivative if the light curve is concave or convex.
Details in the accretion geometry might make the sign of the
correlation positive or negative for near-polar hot spots, but this
remains to be calculated.  The observed X-ray flux threshold below
which the pulse phase deviates from the linear correlation (as in
J1814, J0929, J1808, and J1807) might indicate the onset of a
different mode of accretion (such as a propeller).

True underlying spin up or spin down episodes during an outburst
cannot be excluded from these observations.  However, our analyses
indicate that if phase depends on flux as we suggest any true {{\it spin}}
frequency derivative $\dot\nu_s$ must be much lower than previously
claimed, as the measured $\dot\nu$ values primarily result from the
X-ray lightcurve concavity. We then interpret the best fit reference
frequencies derived under the hypothesis of a uniform phase-flux
correlation (Table~1) as our best estimates of the true spin frequency
in each outburst.  
With our new set
of spin frequencies for J1808 we infer a long term spindown which,
interpreted as due to the magnetic dipole torque, implies a dipole
moment of $\mu=1.4\pm0.1\times 10^{26}\rm\,G\,cm^3$ (see \citealt{s06}
for a derivation of the non-vacuum magnetic dipole torque) and a
magnetic field strength at the poles in the range
$B=2.0$--$2.8\pm0.2\times 10^8\rm\,G$
This value is slightly smaller than the previous estimate of
H08 and agrees with that inferred from optical and quiescent
X-ray observations ($B\approx 2-10\times 10^{8} \rm\,G$,
\citealt{b03}, \citealt{db03}) and with the value expected from
standard accretion theory (\citealt{wvdk98}, \citealt{pc99}).

An open question is why in J1808 in 1998 and
in J1807 the correlation is different for overtone and
fundamental.  A difference in behavior between fundamental and
overtone was first noted by \citet{b06} for the 2002 outburst of
J1808, who suggested that due to competing contributions from two
poles to the pulse profile the overtone more closely tracks the spin.
However, if, as our results suggest, the timing noise in the
fundamental can be explained from the flux variations through a moving
hot spot model, then after correcting for this the fundamental must
reflect the spin.  We note that, {\it e.g.}, an $\dot M$ dependent
competition between fan and pencil beam contributions to the pulse
profile would instead primarily affect the phase of the
overtone. 

In conclusion, our analysis of a large record of AMXP data suggests
that $\dot{M}$ induced hot spot motion dominates the observed pulse
phase residual variations and that this effect needs to be taken into
account when measuring the spin of these neutron stars.


\begin{thebibliography}{99}
\expandafter\ifx\csname natexlab\endcsname\relax\def\natexlab#1{#1}\fi
\expandafter\ifx\csname url\endcsname\relax
  \def\url#1{{\tt #1}}\fi
\expandafter\ifx\csname urlprefix\endcsname\relax\def\urlprefix{URL }\fi

\bibitem[{{Bildsten} et~al.(1997){Bildsten}, {Chakrabarty}, {Chiu}
  et~al.}]{b97}
{Bildsten} L., {Chakrabarty} D., et~al., Dec. 1997, \apjs, 113, 367

\bibitem[{{Burderi} et~al.(2003){Burderi}, {Di Salvo}, {D'Antona}, {Robba}, \&
  {Testa}}]{b03}
{Burderi} L., {Di Salvo} T., et~al., Jun. 2003, \aap, 404, L43

\bibitem[{{Burderi} et~al.(2006){Burderi}, {Di Salvo}, {Menna}, {Riggio}, \&
  {Papitto}}]{b06}
{Burderi} L., {Di Salvo} T., et~al., Dec. 2006, \apjl, 653, L133

\bibitem[{{Chou} et~al.(2008){Chou}, {Chung}, {Hu}, \& {Yang}}]{c08}
{Chou} Y., {Chung} Y., et~al., May 2008, \apj, 678, 1316

\bibitem[{{Di Salvo} \& {Burderi}(2003)}]{db03}
{Di Salvo} T., {Burderi} L., Jan. 2003, \aap, 397, 723

\bibitem[{{di Salvo} et~al.(2007){di Salvo}, {Burderi}, {Riggio}, {Papitto}, \&
  {Menna}}]{d07}
{di Salvo} T., {Burderi} L., et~al., Aug. 2007, In: {di Salvo} T., {Israel}
  G.L., et~al. (eds.) The Multicolored Landscape of Compact Objects and Their
  Explosive Origins, vol. 924 of American Institute of Physics Conference
  Series, 613--622

\bibitem[{{Falanga} et~al.(2005){Falanga}, {Kuiper}, {Poutanen} et~al.}]{f05}
{Falanga} M., {Kuiper} L., et~al., Dec. 2005, \aap, 444, 15

\bibitem[{{Galloway} et~al.(2002){Galloway}, {Chakrabarty}, {Morgan}, \&
  {Remillard}}]{g02}
{Galloway} D.K., {Chakrabarty} D., et~al., Sep. 2002, \apjl, 576, L137

\bibitem[{{Galloway} et~al.(2007){Galloway}, {Morgan}, {Krauss}, {Kaaret}, \&
  {Chakrabarty}}]{g07}
{Galloway} D.K., {Morgan} E.H., et~al., Jan. 2007, \apjl, 654, L73

\bibitem[{{Hartman} et~al.(2008){Hartman}, {Patruno}, {Chakrabarty}
  et~al.}]{h08}
{Hartman} J.M., {Patruno} A., et~al., Mar. 2008, \apj, 675, 1468

\bibitem[{{Jahoda} et~al.(2006){Jahoda}, {Markwardt}, {Radeva} et~al.}]{j06}
{Jahoda} K., {Markwardt} C.B., et~al., Apr. 2006, \apjs, 163, 401

\bibitem[Lamb et al.(2008)]{l08} Lamb, F.~K., Boutloukos, 
S., Van Wassenhove, S., Chamberlain, R.~T., Lo, K.~H., Clare, A., Yu, W., 
\& Miller, M.~C.\ 2008, arXiv:0808.4159 

\bibitem[{{Papitto} et~al.(2007){Papitto}, {di Salvo}, {Burderi}
  et~al.}]{pap07}
{Papitto} A., {di Salvo} T., et~al., Mar. 2007, \mnras, 375, 971

\bibitem[Patruno(2008)]{pat08} Patruno, A.\ 2008, American 
Institute of Physics Conference Series, 1068, 25 

\bibitem[Patruno et al.(2009)]{pat09} Patruno, A., 
Altamirano, D., Hessels, J.~W.~T., Casella, P., Wijnands, R., 
\& van der Klis, M.\ 2009, \apj, 690, 1856 

\bibitem[Poutanen(2006)]{pou06} Poutanen, J.\ 2006, Advances 
in Space Research, 38, 2697 

\bibitem[{{Poutanen} \& {Gierli{\'n}ski}(2003)}]{pg03}
{Poutanen} J., {Gierli{\'n}ski} M., Aug. 2003, \mnras, 343, 1301

\bibitem[{{Psaltis} \& {Chakrabarty}(1999)}]{pc99}
{Psaltis} D., {Chakrabarty} D., Aug. 1999, \apj, 521, 332

\bibitem[{{Riggio} et~al.(2008){Riggio}, {Di Salvo}, {Burderi} et~al.}]{r08}
{Riggio} A., {Di Salvo} T., et~al., May 2008, \apj, 678, 1273

\bibitem[Romanova et al.(2003)]{r03} Romanova, M.~M., 
Ustyugova, G.~V., Koldoba, A.~V., Wick, J.~V., 
\& Lovelace, R.~V.~E.\ 2003, \apj, 595, 1

\bibitem[{{Romanova} et~al.(2004){Romanova}, {Ustyugova}, {Koldoba}, \&
  {Lovelace}}]{r04}
{Romanova} M.M., {Ustyugova} G.V., et~al., Dec. 2004, \apjl, 616, L151

\bibitem[{{Spitkovsky}(2006)}]{s06}
{Spitkovsky} A., Sep. 2006, \apjl, 648, L51

\bibitem[{{Watts} et~al.(2008){Watts}, {Patruno}, \& {van der Klis}}]{w08}
{Watts} A.L., {Patruno} A., {van der Klis} M., Nov. 2008, \apjl, 688, L37

\bibitem[{{Wijnands}(2004)}]{w04}
{Wijnands} R., Jun. 2004, Nuclear Physics B Proceedings Supplements, 132, 496

\bibitem[Wijnands \& van der Klis(1998)]{wvdk98} Wijnands, R., \& van der Klis, M.\ 1998, \nat, 394, 344 
\end{thebibliography}

\end{document}